# Me Love (SYN-)Cookies: SYN Flood Mitigation in Programmable Data Planes

Dominik Scholz, Sebastian Gallenmüller, Henning Stubbe, Bassam Jaber, Minoo Rouhi, Georg Carle
*Technical University of Munich*,
{scholz, gallenmu, stubbe, rouhi, carle}@net.in.tum.de, bassam.jaber@tum.de

*Abstract*—The SYN flood attack is a common attack strategy on the Internet, which tries to overload services with requests leading to a Denial-of-Service (DoS). Highly asymmetric costs for connection setup—putting the main burden on the attackee—make SYN flooding an efficient and popular DoS attack strategy. Abusing the widely used TCP as an attack vector complicates the detection of malicious traffic and its prevention utilizing naive connection blocking strategies.

Modern programmable data plane devices are capable of handling traffic in the 10 Gbit/s range without overloading. We discuss how we can harness their performance to defend entire networks against SYN flood attacks. Therefore, we analyze different defense strategies, SYN authentication and SYN cookie, and discuss implementation difficulties when ported to different target data planes: software, network processors, and FPGAs. We provide prototype implementations and performance figures for all three platforms. Further, we fully disclose the artifacts leading to the experiments described in this work.

## I. INTRODUCTION

A TCP connection setup requires a three-way handshake initiated by the client. First, the client sends a SYN segment, selecting a random sequence number and setting its TCP options. The server has to respond accordingly while storing the client's sequence number along additional connection-specific metadata in a *backlog* containing Transmission Control Block (TCB) entries. This server-side state allocation has to be done for each SYN segment received, which is the weakness exploited by the SYN flood attack. By sending SYN segments with different (spoofed) source addresses and ports to the server, the capacity of the *backlog* is exhausted [17]. If all entries of the *backlog* are allocated, newly arriving SYN segments will either be dropped, or old entries will be reused [50]. Consequentially, SYN segments of a legitimate client will either be dropped, or state might be evicted before the client responds with the final segment of the handshake. Thus, attacks lead to a DoS, as no new connections can be opened.

In April of 2013 Israeli cyberspace was attacked, whereby SYN flooding was a primary attack vector with up to 82 kpps, exhausting both firewall and web server resources [41]. However, it was not the only attack vector, a UDP flood, saturating the network's bandwidth, as well as DNS and ICMP floods were utilized. Furthermore, a RST flood was launched to additionally cripple web server services. Similar behavior was observed during an attack in September of 2013 targeting large banking groups [42]. A SYN flood was part of a larger UDP and HTTP flood, while specialized tools exploited web server vulnerabilities. According to Kaspersky Lab's quarterly reports [33], from 2017 to 2019 the share of SYN flood traffic during large-scale DDoS attacks rose up to 84 % becoming the "most popular type of attack" [33] traffic.

Common mitigation strategies—SYN cookie and SYN authentication—require the client to behave correctly beyond the initial SYN segment. The strength of the defense capabilities relies on the performance available to enforce and check correct TCP behavior before initiating a costly TCP connection. In this work we want to investigate how powerful off-the-shelf data planes such as the software-based DPDK, or programmable hardware such as Network Processing Units (NPU) or FPGAs can be used for SYN flood defense.

We want to answer, *(i)* why endhosts cannot effectively defend themselves against SYN flood attacks, *(ii)* what different available defense strategies and their impact on TCP are, and finally *(iii)* how we can leverage the power of modern programmable data planes to provide flexible, scalable and extendable solutions to mitigate SYN flood attacks.

The paper is outlined as follows. Section II summarizes the state of the art for SYN flood defense and highlights how SYN flood defense is currently implemented in Linux. Based on this, we introduce our concept for a scalable SYN proxy solution in Section III. Sections IV and V present our prototype implementations using two different approaches: the libmoon/DPDK software packet processing framework and using the P4 domain specific language (DSL). We compare our prototype implementations in regard to performance and resource consumption metrics in Section VI. Related work is presented in Section VII. Section VIII concludes the paper.

## II. STATE OF THE ART

Since the first mention of SYN flood attacks, manifold proposals how to mitigate them have surfaced and implemented in network stacks. In general, the mitigation methods can either be applied on the end-host or handled externally.

### A. Mitigation at Node Level

As the attack targets the TCP/IP stack of the server, **parameter optimizations** of the stack can be utilized on the target server itself. Tweaking parameters or architecture of the application results in only minimal, if at all, any success in throttling the attack as these adjustments only affect the overall performance of the TCP/IP stack and application [48, 6].

**SYN flood-specific parameter adjustments** of the TCP/IP stack include increasing the *backlog* or reducing the SYN/ACK retransmission times, thereby reducing the timeout per TCB. Considering today's bandwidths, modifying these parameters only marginally delays the problem by using more memory resources. In fact, the *backlog* structure might not be designed to surpass a certain size, as data structures or search algorithms used might become inefficient. Thus, such modifications may worsen performance without an attack [16]. Similar, reducing timeouts might lead to legitimate connections, with a large RTT, not being fast enough to finish the handshake before the entry gets deleted [16].

**Replacing TCB entries** either randomly or by choosing the oldest entry is another approach to modify the *backlog*. While this works for SYN flood rates of up to approx. 500 SYN segments per second [44], for packet rates possible in today's networks legitimate connections are evicted with high probability ($> 99\%$) before they get established.

Instead of storing the full TCB, only absolutely required data, which can be as few as 16 B [27], can be stored in the **SYN cache** approach. Secret bits of the SYN segment, in combination with the addresses and ports, are hashed, to determine the bucket in a hash map [16, 27, 17, 36]. In FreeBSD the size of such an entry is reduced to 160 B, compared to 736 B for a TCB [17] in Linux, and can contain more than 15 000 entries. Furthermore, the linked-list of each bucket has a limited size, i.e., if a bucket would overflow, the oldest entry gets deleted [35]. The idea is that the attacker does not know how the hash is calculated, hence, can't attack the hash structure. Appropriate hash function choice ensures that memory and computational resources required are limited [16]. A downside is that TCP options have restricted or no support. The attack will eventually be successful when surpassing a high enough SYN flood rate to exhaust the data structure.

A widely deployed method are **SYN cookies** [9]. State, usually kept by the server, is instead encoded and put into the sequence number $n_{syn}$, which can be freely chosen. A legitimate client will finish the handshake sending a TCP ACK segment, setting the acknowledgment number to $n_{syn} + 1$. The server only accepts the connection and creates necessary state, if the received number can be decoded and verified.

While using SYN cookies, the 32 bit sequence number is made up of three different values. Five bits store a timestamp value, that has an accuracy of 64 s, to prevent the collection and re-injection of old cookies [9]. Three bit encode the Maximum Segment Size option as it is essential for TCP performance. The remaining 24 bit are used to store a cryptographic hash of the connection 4-tuple (source and destination IP addresses and ports; as the attack is exclusive to TCP the IP protocol field is not hashed) and the timestamp value. Connection state is only created if the timestamp is not outdated and the hash is correct.

SYN cookies perform a tradeoff: instead of consuming memory resources (connection depletion attack), CPU resources are exhausted [28]. Even an attack with low bandwidth can shut down a high-performance server, exploiting the asymmetric nature of the attack [30]. The server not only has to calculate the cookie, but also verify it in the received ACK segment. In theory, this makes the SYN cookie approach susceptible to ACK floods. However, the computational effort is the same as during a SYN flood calculating the cookie.

Similar to the SYN cache, a drawback of this approach is the restricted support of TCP options. An updated SYN cookie layout uses the TCP timestamp field [17] to also encode other options that are widely used today [40].

The goal of **SYN authentication** is to not have the server invest resources to verify the clients legitimacy. Instead, the server expects a certain response to a triggered, unusual event, which can be compared to client puzzles [8, 28, 52]. If the client reacts accordingly, any further connection attempt is whitelisted. Multiple approaches for challenges exist. The simplest idea is to ignore or reset the initial connection attempt, however, an attacker can simply resend the SYN segment and is whitelisted. More advanced, the server sends, e.g., a SYN/ACK segment with invalid sequence number. Clients are expected to respond with a RST segment [27] (referred to as $\text{Auth}_{\text{wrong}}$). Even so, an attacker can circumvent this, by including a flood of RST segments in their attack. Other SYN authentication approaches only whitelist once the client has demonstrated its intent to finish a complete handshake (final TCP ACK segment is received, referred to as $\text{Auth}_{\text{full}}$) or even by means of a higher layer protocol (e.g. a HTTP GET request is received). Fingerprinting techniques can be used to increase the probability that SYN and ACK/RST segments for a handshake were sent by the same client (spoof protection). For instance, the IPv4 TTL [39] (referred to as $\text{Auth}_{\text{TTL}}$) or TCP options signature can be compared. All SYN authentication approaches can be further enhanced by including a cookie value as sequence number, which can then be verified on reception of the RST [44] or the ACK (referred to as $\text{Auth}_{\text{cookie}}$). Similar to SYN cookies, the hash calculation can be exploited by an attacker.

During the initial connection attempt no TCB state is created. Instead, only when the client reacts accordingly, it gets whitelisted, which can be accomplished by storing one bit per flow, source address or source subnet. In particular, no information about TCP options (or only a signature when fingerprinting) is stored. Successive connection attempts do not have to be modified or inspected beyond the SYN flag, until the whitelist entry is invalidated.

A drawback is the interruption of the regular protocol flow through a reset, increasing the delay for the initial connection. Furthermore, if the attacker manages to get whitelisted, a subsequent SYN flood is considered legitimate traffic.

All of the mentioned mitigation strategies can be combined to form **hybrid approaches**. Aside from tweaking parameters for optimization, more advanced strategies like the SYN cache, SYN cookies or SYN authentication are usually only activated on demand when an attack is detected [17].

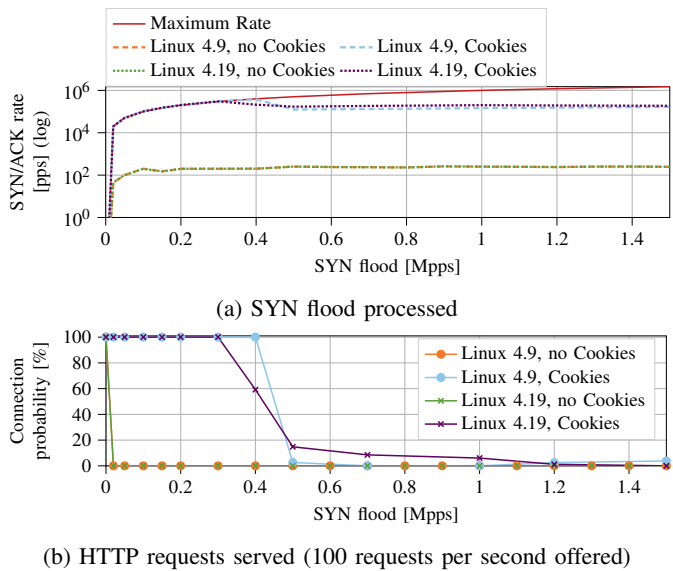

(a) SYN flood processed

(b) HTTP requests served (100 requests per second offered)

Fig. 1: Performance of Linux during SYN flood

### B. Case Study: SYN Cookies in Linux

Linux's TCP/IP network stack has SYN cookies enabled by default, but only uses them when the backlog of a socket is already full [3]. Linux follows the proposed SYN cookie layout, encoding the MSS and a timestamp value with 60 s resolution. If the client sent a timestamp option it is used to encode further TCP options. As a case study, we analyze two different Linux versions when subjected to a SYN flood, while serving a static website (see Figure 6a): 4.9.0 and 4.19.0, using SHA1 and SipHash as hash function, respectively.

Figure 1 shows the amount of processed SYN flood as well as the probability to successfully serve 100 HTTP requests per second for an increasing SYN flood. When disabling SYN cookies, Linux can only process up to 250 SYN packets per second. However, no HTTP requests are served. Profiling (see Figure 2a) reveals that this behavior is not due to CPU exhaustion. Instead, the backlog size is the limiting factor.

When enabling SYN cookies both Linux versions behave similar, i.e., process up to 0.4 Mpps of SYN flood. Moreover, up to this point, all HTTP requests are served. When further increasing the SYN flood, the probability of serving any legitimate requests successfully approaches zero. For Linux 4.19 it is notable that a small percentage ($< 10\%$) of requests is served, even for rates of up to 1 Mpps SYN flood.

Profiling (Figure 2b and Figure 2c) shows that, in this setting, CPU utilization is the limiting factor. A clear difference between the Linux versions is the amount of CPU cycles used for hash calculation. Cookie calculation using SHA1 requires up to 15 % of the cycle budget, while SipHash (2.5 %) is more efficient. Given the cheaper hash function, one would expect an increased performance for Linux 4.19 instead of the measured decrease. We attribute this development to other networking related changes between these Linux versions, which are out of scope of this study. Profiling also reveals common performance limiting factors of the Linux network stack. The majority of CPU cycles are spent handling packet buffer and memory (skb/mem), driver related processing (ixgbe/dev), IP layer processing (ip/inet), and the TCP stack itself (tcp). The numbers are in the expected range and comparable to other Linux studies [43].

Primary goal of the network stack is to support as many protocols as possible, providing a reliable, stable and robust interface for userspace applications. Thus, raw performance is secondary, wherefore Linux should not be used for mitigation of large-scale SYN floods. Therefore, optimized solutions are required. One approach to increase performance is to perform filtering operations like DoS mitigation before the network stack, e.g. by using the eBPF-programmable XDP [47].

Mere parameter changes to the TCP stack cannot effectively defend against a SYN flood. Out of the more promising techniques, SYN authentication and SYN cookie, Linux supports the latter. However, our measurements show that Linux, in the best case scenario, can cope with a maximum attack rate of 0.4 Mpps. Moreover, this strategy protects only applications running on this node, i.e., in a cluster of nodes, each node has to perform its own SYN flood mitigation, wasting CPU cycles that could otherwise be used for running applications.

### C. Mitigation at Network Level

SYN flooding is most successful when utilizing spoofed source addresses. Therefore, applying **ingress filtering** [20] would reduce effectiveness of this and other DoS attacks relying on spoofed addresses. Due to the distributed characteristic of the Internet, this, however, is unlikely to happen [36].

Active or passive **traffic monitoring** of the total traffic in the network can be used to detect anomalies, specifically to detect higher than usual volumes of SYN segments, at the network's edge [17]. Machine learning or other means (e.g. CUSUM [12, 13]) can be used to discern abnormal traffic patterns. As result, the attack can be thwarted or throttled before the traffic reaches the targeted server.

SYN flood mitigation strategies can be deployed either as part of the TCP/IP stack on the server itself, or as separate host, acting as **SYN proxy**. Advantages of using a dedicated proxy include that the server's resources are not utilized for mitigation, as well as, that more than one server can be protected. The concept of a SYN proxy is based on the idea of intercepting potentially harmful traffic before it reaches the server. Instead of the server, the proxy will answer to the initial SYN segment [17, 36, 16] using for instance SYN cookies or SYN authentication and only forward legitimate packets to the server. Proxy services are often times combined with monitoring of the traffic, i.e., only when an attack is detected, traffic gets re-routed through the proxy.

A second approach for a SYN proxy is to allow all SYN segments to reach the server as usual. However, for every SYN/ACK that the server sends, the proxy will immediately reply with a spoofed TCP ACK, which finalizes the connection, at least from the point of view of the server, and frees *backlog* space occupied by the TCB [17]. If the client is legitimate and sends an ACK segment, the server will simply ignore it as it was already received before. To cope with malicious SYNs, the proxy will send a RST to the server after a certain timeout to terminate the tentatively created connection. The advantage of this strategy is that the proxy does not "actively participate in legitimate connections after they are established" [17]. However, the server is still subject to the SYN flood and also has to handle a large amount of established TCP connections.

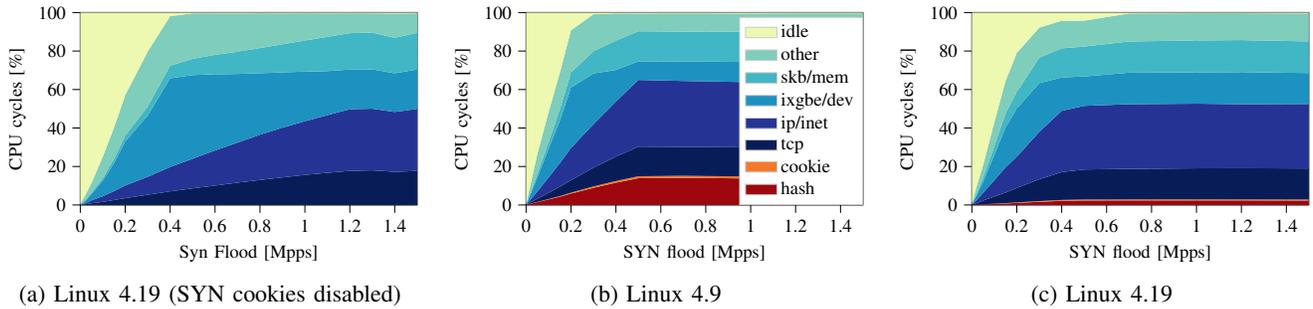

Fig. 2: Profiling of Linux during SYN flood

### D. Comparison

In Table I we compare the promising approaches based on metrics also discussed in literature [44]. The weakness of the SYN cache is the memory exhaustion, while SYN cookies shift to utilizing CPU resources. SYN authentication strategies do neither require extensive memory nor CPU ressources. The exception are $Auth_{TTL}$ to store TTL values, and $Auth_{cookie}$ to perform cryptographic hashes. Calculating cookie values is also the limiting factor for the efficiency, i.e., how much SYN flood can be processed. None of the SYN authentication strategies is transparent for the client application, as they reset the initial connection. It has to be noted, that for instance for modern browser implementations an automatic retry is performed in this case, resulting in no significant interruption for the user. A downside of SYN cookies is the limited support for TCP options. Robustness specifies whether the signaling capabilities of TCP are compromised [44]. This is the case for all strategies besides the SYN cache, as no initial state is kept, wherefore a retransmission of a SYN/ACK segment is not possible. This is more severe for SYN cookies than for SYN authentication, as the latter work on the assumption that the client retries failed connection attempts anyway.

Lastly, we evaluate the techniques by whether they correctly classify legitimate traffic. Assuming all packets are received correctly and within the timeframe of the calculated cookie, no technique produces false negatives (legitimate traffic classified as malicious traffic). Only the SYN cache, SYN cookies and $Auth_{cookie}$ cannot be circumvented by malicious traffic, i.e., only a legitimate connection is passed to the server. No other SYN authentication strategy fulfills this requirement, as a RST or ACK flood can achieve whitelisting for malicious SYN segments as no cryptographic cookie is used.

|  | SYN | | Authentication | | |
|---|---|---|---|---|---|
|  | Cache | Cookie | Wrong | Full | TTL | Cookie |
| Mem. immunity | – | + | + | + | - | + |
| CPU immunity | + | - | + | + | + | - |
| Efficiency | - | + | ++ | ++ | ++ | + |
| Transparency | + | + | - | - | - | - |
| Option support | + | -/o | + | + | + | + |
| Robustness | + | - | o | o | o | o |
| Classified as legitimate traffic | | | | | | |
| False Positive | + | + | - | - | o | + |
| False Negative | + | + | + | + | + | + |

TABLE I: Comparison of SYN flood mitigation strategies

It has to be noted that these techniques are usually only deployed during an ongoing attack. Therefore, the assumption is that the server is under severe load caused by a SYN flood, while only a small percentage of the traffic is legitimate TCP traffic. The goal is to provide a best-effort approach of service quality for legitimate flows, i.e., higher delay or minor connection disruptions are acceptable, as long as any traffic is served by the webserver.

None of the listed methods reliably shuts down the attack, instead the success of the attack is delayed by having to use more resources for the attack. Furthermore, no strategy is part of the official TCP specification or standardized by a committee [17]. The different strategies offer a tradeoff between efficiency (performance) and correct classification (false positives). For the remainder of the paper we focus on SYN cookies and SYN authentication performing a full handshake and using a cookie, as only these strategies offer good protection. For the latter, we also discuss not using the cookie to highlight the impact of performing the cryptographic hash.

## III. SYN PROXY CONCEPT

We identified a SYN proxy (see Figure 3) being the most efficient and effective way of protecting a multitude of servers without using their resources. We discuss the requirements for implementing a flexible and open-source SYN proxy and highlight challenges encountered when using different mitigation strategies in this setup.

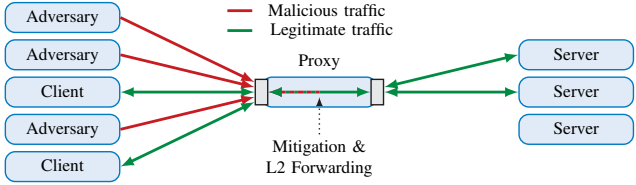

Fig. 3: SYN proxy scenario

## A. Requirements

We have identified the following requirements for a high-performance SYN proxy solution:

**SYN Flood Mitigation:** No malicious SYN flood traffic should reach the protected server(s). A proxy implementation should filter and drop malicious traffic while detecting and forwarding legitimate connection attempts.

**Performance:** The proxy should be able to operate at line rate of $10\,\text{GbE}$, i.e., it should be able to process $14.88\,\text{Mpps}$ of SYN flood. Furthermore, it should be scalable to even higher data rates (e.g. $100\,\text{GbE}$).

**Service Quality:** Services provided by proxied servers should be reachable for as many legitimate clients as possible. As connections might occur during a DoS attack period, a best effort approach regarding service quality for legitimate traffic is chosen.

Furthermore, the behavior of the proxy should adhere to standards, whenever possible, to guarantee the best possible interaction with different client and server implementations.

**Portability:** To ease adoption, the proxy implementation should focus on platform independence, facilitating deployments on a wide range of different hardware. Additionally, fine-grained configurability should be possible, preventing setup dependencies, whenever possible.

**Extensibility:** The proxy should serve as starting point that allows for further SYN flood mitigation strategies, and mitigation of other types DoS attacks, e.g. UDP or DNS floods.

While the proxy should perform well without any malicious traffic, the concept of a SYN flood proxy, or in general an attack mitigating proxy, includes, that it is only deployed during an attack. Therefore, the proxy is designed with the assumption that more than $90\,\%$ of all received traffic is malicious, i.e., only a minority of traffic is considered legitimate.

## B. Mitigation Strategies

Based on the overview of mitigation strategies presented in Section II we choose SYN cookies and different SYN authentication variants for our prototype implementations.

*1) SYN Cookies:* SYN cookies are the first choice for SYN flood mitigation as they are transparent to client and server, i.e., no protocol violations or connection resets, but only offer limited support for important TCP options.

However, when used in a proxy setup, SYN cookies raise several issues. Once client and proxy have finished the initial handshake (see Figure 4), the proxy cannot simply forward segments of the client to the server, as the server is still unaware of the connection and has no state for it. To uphold transparency of SYN cookies, the proxy has to start a second connection between itself and the server. While the proxy can reuse the first handshake's sequence number in its SYN segment (seq = x), the server will choose its sequence number at random (z), i.e., it does not match the proxy's sequence number of the connection with the client (y). Consequently, the proxy has to calculate and store the difference between these sequence numbers. Later on, the proxy has to modify sequence and acknowledgment numbers of every future segment of the connection accordingly.

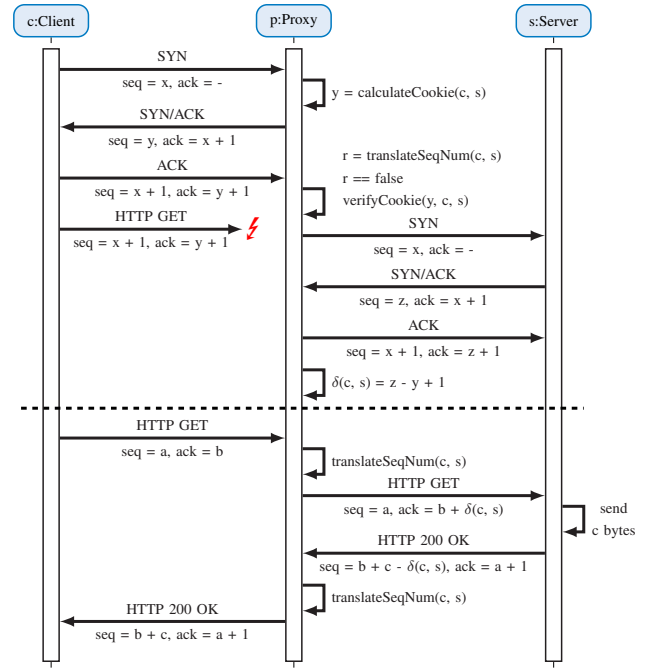

Fig. 4: Message exchange for the SYN cookie strategy

The second issue is that the first segment of data by the client is sent directly after the third message of the first handshake (see Figure 4). This poses a problem for the SYN proxy, as the second handshake with the server is not completed yet. Hence, the proxy cannot yet translate and forward the data segment. As a consequence, the client retransmits this segment after a timeout, for which $200\,\text{ms}$ is a common time period. Considering that modern RTTs are only a hundredth of this, waiting for the retransmission is a significant delay penalty and undesirable for the client.

A solution to combat this problem is for the proxy to store the initial data segment instead of dropping it. Once the second handshake is completed, the stored segment can be translated and forwarded. Alternatively, the proxy can actively notify the client once the second handshake is complete, by resending the SYN/ACK segment, triggering a retransmission of the data segment. Further improvements include setting a window size of zero (zero window) in the original SYN/ACK segment, indicating that the server cannot process any more data. Until the SYN/ACK is resent with a non-zero window size, the client will not send the initial data segment, reducing bandwidth wasted for a segment that will be dropped.

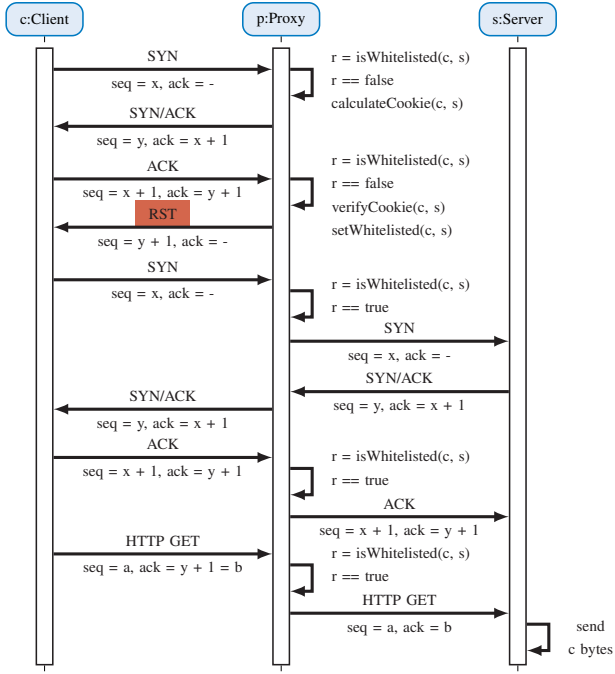

Fig. 5: Message exchange for the Auth$_{cookie}$ strategy

|  | SYN Cookies | SYN Authentication |
|---|---|---|
| Packet modification | every segment | handshake |
| Transparent | yes | no |
| Option support | limited (encoded) | full |
| State |  |  |
| State per | flow | flow/subnet |
| Lookup key | 4-tuple | 4-tuple/src. IP |
| Memory required | > 32 bit | 2 bit |
| Lookup for | not SYN | every segment |

TABLE II: Comparison of SYN cookie and SYN authentication in proxy setup

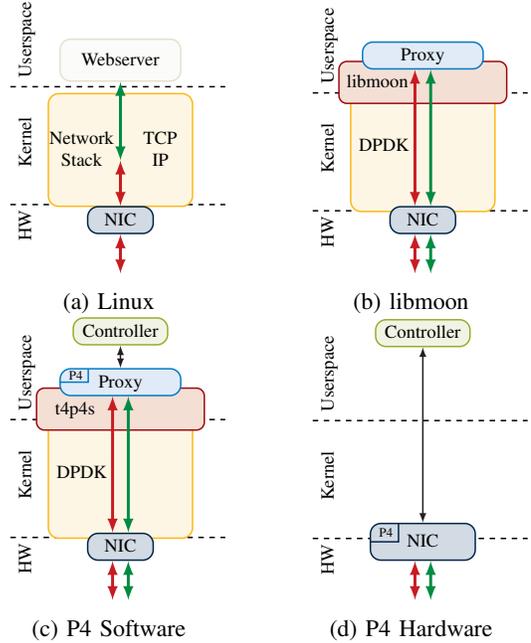

Fig. 6: Architecture of SYN flood mitigation mechanism when used in Linux or as SYN proxy using software packet processing or P4.

*2) SYN Authentication:* SYN authentication is the most efficient solution due to its simplicity. As depicted in Figure 5, the first connection attempt is interrupted. By whitelisting all further attempts for this client, the proxy does not have to keep a separate connection with the server or perform sequence number translation. No connection state has to be stored by the proxy, unless using Auth$_{TTL}$, wherefore a simple bitmap representing the whitelist is sufficient.

However, as ACK segments of established connections cannot be distinguished from the third segment of the handshake, the proxy has to check every segment against the whitelist. If the origin is not whitelisted, the segment is assumed to be the third segment of the handshake and, when using Auth$_{cookie}$, the cookie hash is verified.

We focus on Auth$_{full}$ and Auth$_{cookie}$ to analyze the performance impact of calculating cookie values.

*3) Comparison:* Both strategies implemented as proxy have differences in regards to packets modified, transparency and option support (see Table II). For SYN authentication state size depends on the whitelisting granularity, e.g. per flow or for full subnets. However, state is reduced to one or two bit per entry compared to keeping the sequence number difference per flow for SYN cookies.

## IV. SOFTWARE PACKET PROCESSING PROTOTYPE

We use libmoon [23, 18] to implement our SYN proxy prototype in software targeting commercial off-the-shelf (COTS) hardware. libmoon offers a powerful and easy-to-use scripting layer on top of DPDK's [2] raw packet handling (see Figure 6b). As libmoon offers utility for header and packet operations for a wide variety of protocols, crafting new and individual packets is possible while retaining the performance of DPDK [23]. The proxy is running as userspace program, bypassing the regular kernel network stack. libmoon does not offer a TCP stack therefore the handling of the TCP handshake has to be done by the proxy application itself. Utility functions for packet templating are available to simplify the implementation of the TCP handshake. libmoon receives and processes packets as batches. For each packet of a batch, depending on the mitigation strategy and TCP flags set, a response is crafted and sent after all packets of the batch have been processed.

### A. Hash Function

We use the pseudo-cryptographic SipHash [7] function to calculate hash values for cookies and hashmap keys. It is optimized for performance when using short inputs and is used in many applications, including DoS protection [7]. It has been shown that SipHash can be integrated efficiently with programmable software and hardware data planes [46].

## B. Connection State Tracking

A crucial part of state maintenance is correct tracking of TCP connection state, i.e., adding new connections and to regularly remove dead connections. Removing state of finished connections serves two purposes: freeing memory space and allowing the same 4-tuple to initiate a new connection. Approaches to state removal are either active, or passive. **State tracking** aims at actively reacting to FIN and RST segments of the connection to determine when the connection concludes. For each flow, state kept by the proxy has to reflect whether such a segment has been seen. If state tracking concludes that the connection has been terminated, i.e., both sides sent and acknowledged a FIN segment or a RST has been sent, the corresponding state entry can be removed immediately. This approach is difficult to implement as reordering of segments, packet loss, retransmissions or connection timeouts, may result in state being deleted too early, or not at all. A passive approach is to use garbage collection, i.e., applying a **second-chance** page replacement algorithm [14]. For this, each state entry is extended with two bits, which are set every time the entry is looked up. A periodic background process iterates all entries and unsets one of the bits. If both bits are unset, the entry is deemed inactive and removed. Common timeouts for inactive connections for instance in Linux are at least 10 min [3].

## C. SYN Authentication

SYN authentication requires no state information besides the previously described connection state tracking. Therefore, we choose a fixed-size bitmap as data-structure representing the whitelist. For each entry, two bits are used to implement the second-chance page replacement algorithm, resulting in a total of approx. 1 GB of memory being used when representing the complete IPv4 address space and whitelisting based on individual source addresses. As the location of the bits for a particular entry is known, no hash function to determine the key has to be used. For IPv6, either whitelisting based on subnets or a hash-based data-structure has to be used.

## D. SYN Cookies

The SYN cookie approach requires additional state information beside the connection state tracking. Increased memory consumption make the bitmap impractical. Instead this approach uses concurrent hash maps, as the garbage collection process has to run as separate process to not interfere and block time-sensitive lookups. When operating a normal hash map implementation, every insert operation would invalidate the iterator object used to traverse the map. Furthermore, when using only this mechanism to remove invalid entries, re-using of ports is infeasible due to the long timeout.

To leverage the advantages of both approaches—immediate removal of connections and garbage collection of left-over entries—we use a **hybrid approach** coined swap-out hash maps. Instead of a concurrent hash map and timeout bits, two normal hash maps (`active` and `history`) are used in conjunction. Rather than periodically unsetting bits, `history` is deleted and replaced by the `active` map, and a new empty `active` hash map is initialized.

The lookup procedure consists of two stages. First, the `active` map is checked. If the entry exists, it is returned and no further action is performed. If not, the same lookup is performed in the `history` map. If no entry is found, the connection is assumed to not exist. Otherwise, the entry is copied to the `active` map, employing the second chance mechanism, and returned. Inactive entries are eventually swapped out to the `history` map before being deleted with the next swap. Insert operations are performed exclusively on the `active` hash map. With this scheme, no hash map is traversed, removing the need of a concurrent hash map.

Because of the primary assumption of the proxy, i.e., only a margin of all traffic belongs to verified connections, merely a small number of entries is expected to be maintained in the hash maps. Based on this conclusion, we choose Google's sparse hash map implementation [4]. As each value object consists of approximately 40 bit, no bottlenecks regarding required memory space, when keeping thousands of inactive entries up to ten minutes, are expected. The performance impact of a swap-out, and consequent copy-up operations for active connections is reduced by the long timeout interval.

We implemented two strategies to delay the initial TCP data: storing the initial request, and using active notification after completing the handshake with the server. Storing the segment at the proxy requires the complete data to be copied into a buffer, which is stored with the remaining meta data for the connection. This significantly increases complexity and amount of data stored per connection (up to 1500 B), in particular, it requires proper garbage collection of leftover stalled packet buffers to prevent memory leaks. Furthermore, the added complexity quickly makes this approach become infeasible if the client sends more than one data segment before the second handshake is finished or segments are received out of order. For both strategies our evaluation has shown that latency can be reduced to 10 ms or lower, implying that no retransmission after a timeout is required.

## E. Optimizations

To reduce computational complexity, offloading features of modern NICs are used for checksum calculation.

For every TCP segment received by the proxy, exactly one action has to be performed, resulting in one outgoing packet, e.g. sending a SYN/ACK as response to a SYN. Parallelizing the processing by performing all possible actions for each packet of a batch is infeasible as it generates a lot of unwanted, potentially harmful traffic, which would even increase the flooding attack. In all cases, the proxy either answers with a specific pre-crafted (templated) segment, where only addresses etc. have to be updated, or forwards the received segment with slight changes. Reusing RX buffers is only beneficial for the latter case, however, this concerns only a small amount of the overall received traffic, assuming that the majority is SYN flood traffic. Denoting which buffer can be reused and which buffer should be replaced with a templated packet ("dropout batching") would require a bit-mask, which can only be efficiently implemented in C, generating additional overhead per batch [21]. Furthermore, RX buffers can not be reused when being transmitted on a different port.

Instead, RX buffers are never re-used and we utilize on-demand optimistic reduced allocation. For buffers to be forwarded, the complete content is copied, for all other replies we use templated TX buffers from DPDK mempools. For packets that are less likely to be used (RST segments after successful whitelisting) buffers are only allocated on-demand. This reduces overhead when the buffers are not actually used.

## V. Programmable Data Plane Prototype

P4 [11] is a standardized DSL for programming software and hardware data planes. It enables rapid development cycles and creates portable implementations of network applications for ASIC, FPGA, and SmartNIC data planes. Its match-action based paradigm makes it an excellent language to realize packet filtering and DoS mitigation applications.

A difference to using a packet processing framework like libmoon is that P4 does not allow raw packet handling. Packets can only be modified through constructs allowed by the language. Furthermore, P4 has a clear separation between data and control plane: While the data plane is able to match the entries of P4 tables to incoming packets and perform respective actions, only the control plane can insert new entries into a P4 table. The data plane can use digest messages to communicate with the controller, which in our case runs on the same node that also runs the P4 data plane (see Figures 6c and 6d).

### A. Realized Programs and Targets

We have implemented P4 programs for the SYN cookie and SYN authentication strategies and tested their functionality using the Mininet-based bmv2 [1] P4 switch target. For all platforms the P4 program at is core remains the same. However, due to using different P4 architecture models and offering different externs or extern interfaces, all platforms require small modifications to the P4 program. To reduce overhead we only ported the implementations requiring simpler state maintenance, $Auth_{full}$ and $Auth_{cookie}$, to multiple P4 targets: t4p4s [51] (formerly known as P4ELTE) is a DPDK-based P4 software target running on COTS hardware (see Figure 6c). We choose it as it uses the same underlying framework as our libmoon implementation and offers a comparison between implementing raw packet processing to using a DSL for data plane programming. Further, we have ported the implementation to a many-core and a FPGA based hardware data plane, the Agilio Network Flow Processor (NFP)-4000 SmartNIC [5] and the NetFPGA SUME [26], respectively (see Figure 6d).

### B. Program Core

The P4 implementations for the mentioned strategies follow the same structure independent of the used strategy. At first, packets are parsed up to and including the TCP header. The following match-action pipeline at its core functions as a L2 forwarder. Depending on the determined outgoing port, MAC addresses are updated using a table lookup. As P4 cannot generate new packets, the received packet is modified according to the strategy used, set TCP flags, and the state kept by the proxy. State—whitelist or sequence number difference—is maintained as match-action table, requiring one lookup for every segment. No changes to the IP layer are performed besides swapping IP addresses, requiring only an update of the TCP checksum before the packet is transmitted.

### C. Cookie Calculation

To calculate a standard-conform cookie, the P4 target needs to offer functionality for timestamping and hash calculation. Integration of cryptographic hash functions in P4 data planes is possible for software, NFP, and FPGA targets [46]. The integration and use of SipHash as extern on the software target is straight forward as it can be added as library to the hardware dependent t4p4s code. Although the NFP target includes a crypto accelerator for SHA1 and SHA2, it was unavailable on our card, wherefore we opted to integrate a SipHash function as extern. The NFP allows to add extern functions written in a variant of C used to program the processing cores. We are not aware of a P4 ASIC that supports cryptographic functions.

An alternative to raw timestamp access is by using a table containing a counter representing a timestamp value, which is updated by the control plane.

### D. Whitelisting

The easiest approach to perform whitelisting in a P4 program is to use a match-action table. The data plane informs the control plane through a digest message whenever a flow or IP address should be whitelisted and the control plane inserts an according table entry.

An alternative approach that does not include a control plane is to use a bloom-filter data structure built with registers. However, the downsides are the increased complexity and resource consumption of the P4 program, as well as challenges in maintaining the state in the bloom-filter, in particular, evicting outdated entries.

### E. Buffering Packets

Stalling the initial data sent by TCP when using the SYN cookie strategy is not possible when using P4, as P4 has no construct to write entire packets to memory. This is only possible if the target provides an extern for this task, however, it is unlikely due to the memory capacity and memory bandwidth required to perform this operation at line-rate in a programmable ASIC, FPGA, or SmartNIC.

An alternative is to use a secondary COTS device as storage server ("slow-path"), programmed using for instance a framework like libmoon. If a packet needs to be stalled, the proxy forwards this segment to the storage server. Once the handshake is finished, the proxy informs the controller, and the controller instructs the storage server to transmit the stalled packet. The downsides are the increased complexity for the hardware setup, as well as necessary controller logic.

As the underlying problem can be circumvented by using a TCP zero window or active notification, we did not implement this slow-path solution.

## VI. EVALUATION

Software packet processing frameworks allow high flexibility through raw packet processing and easy addition of complex data-structures, but require careful development and optimization, and are limited to software platforms. Using a standardized DSL like P4 makes program development simpler and portable to ASIC, FPGA and SmartNIC devices, but comes at the cost of flexibility as the set of functions offered is limited. The following uses empirical measurements to compare performance metrics of the discussed implementations.

### A. Metrics

The primary performanc indicator for a SYN proxy is the amount of SYN flood processed. From a user perspective, the amount of HTTP requests served without packet loss and the overall latency are a concern. While in general the latency of a device or application when operating in an overload scenario is not of interest, in the case of a SYN proxy it is highly likely that the proxy will reach an overload state during a high volume attack. We therefore analyze latency values in low (no SYN flood), middle (50 % of total processed SYN flood) and overload scenarios.

### B. Measurement Setup

The load generating host sends malicious SYN flood and legitimate HTTP traffic via two separate links to the Device-under-Test (DuT). Using a 10 GbE switch, the traffic is mixed so that malicious and legitimate traffic arrive at the DuT at the same port (i.e. is indistinguishable based on the ingress port). We use Moongen [19] as load generator for the SYN flood traffic. A constant load of HTTP queries is generated using wrk2 [49]. All measurements are run for 30 s, allowing for accurate latency results up to the 99.999 %-ile.

Linux as DuT cannot be configured to run as SYN proxy, therefore, the webserver is located on the same node (see Figure 7a). For all other scenarios using a SYN proxy, the DuT forwards traffic classified as legitimate to a separate host hosting the webserver (see Figure 7b).

For measurements using Linux or t4p4s the DuT is a COTS server system, equipped with an Intel X722 network card and an Intel Xeon Gold 6130 CPU clocked at 2.1 GHz. In other scenarios the DuT is a server equipped with either a 10 GbE NFP-4000 Agilio SmartNIC [5] or a NetFPGA SUME [26].

For all measurements with the COTS system we disabled turbo boost, set the CPU frequency to the maximum of 2.1 GHz, and pinned all traffic to one CPU core. When using Linux as DuT the webserver is pinned to a different CPU core as we are not interested in measuring its performance impact. In all scenarios, the nginx webserver is limited to one worker/CPU and serves a 1 kB static website.

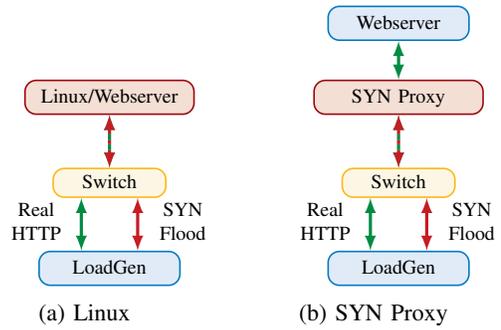

(a) Linux  (b) SYN Proxy

Fig. 7: Measurement setup

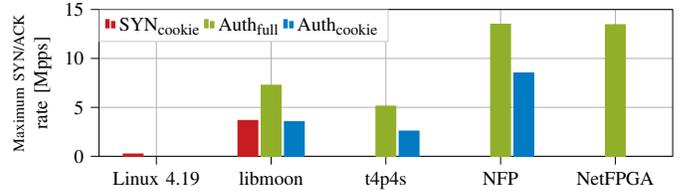

Fig. 8: Maximum processed SYN flood traffic

### C. Webserver Overload

Preliminary tests show that the webserver is capable of processing up to 30000 HTTP requests per second for 10 to 10000 parallel connections. However, latency increases when using more than 1000 parallel connections or more than 4000 requests per second. For more than 700 parallel connections we encountered reduced success probability even when issuing only 100 HTTP requests per second. As we do not want to measure overload artifacts of the webserver node, we restrict our measurements to 100 parallel connections, issuing a total of 100 and 1000 HTTP requests per second.

### D. Processed SYN Flood

Figure 8 shows the maximum amount of SYN flood each implementation is able to process. For all implementations the use of a cryptographic hash function is the limiting factor and reduces the maximum processed SYN flood by up to 50%., which is comparable to other studies [46]. Due to the possibility to optimize and parallelize packet processing, the libmoon implementation achieves up to 50 % better performance than the t4p4s implementation using P4. Only the hardware P4 targets are capable of processing up to 14 Mpps of SYN flood traffic when using the simpler $Auth_{full}$ strategy.

For the software implementations, batch processing has a significant influence on performance (Figure 9). The raw packet handling of libmoon allows the complete processing of packets to be performed in batches. In particular, cookie values are calculated for the complete batch, reducing the overhead of C calls from libmoon. For t4p4s the impact of batching is reduced, as each packet is processed by the P4 pipeline individually, although packets are received in batches. For all remaining measurements we have used the optimal batch sizes of 64 and 32 for libmoon and t4p4s, respectively.

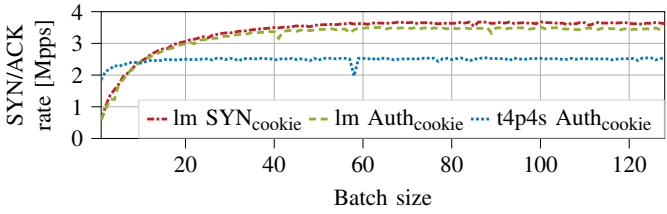

Fig. 9: Processed SYN flood for different batch sizes on COTS system

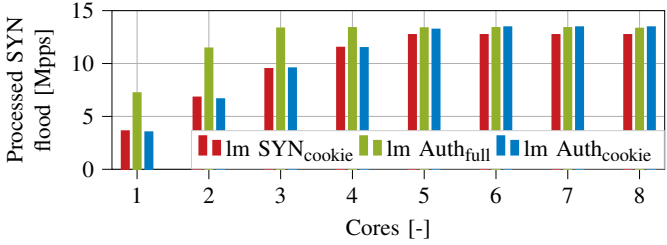

Fig. 10: Multi-core scaling for software system

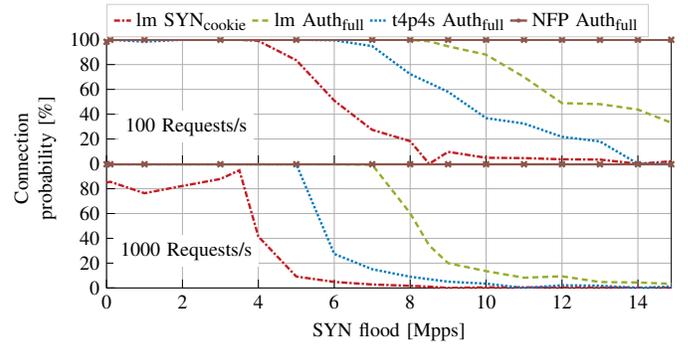

Fig. 11: Success probability for 100 and 1000 HTTP requests per second

The COTS proxy implementations do not have to share state between cores when using receive-side scaling to assign flows to CPU cores. All implemented strategies scale linear with the number of cores (Figure 10), such that even when cookies are calculated, close to line-rate can be reached.

### E. Quality of HTTP Requests

For all implemented solutions the success probability for 100 HTTP requests per second is at 100 % until the point of overload (Figure 11). After this point, the probability slowly drops. As the webserver is not overloaded, all requests reaching the server are served, however, with increasing SYN flood, causing processing overload for the proxy, the chance that the proxy is able to process and forward legitimate traffic drops. As the NFP is able to process the SYN flood at almost line-rate, no packet loss for HTTP packets are encountered.

Using more requests per second reduces the success probability during overload. This is due to the same amount of parallel connections being used. The number of connections experiencing a timeout remain the same, however, in the case of having more requests per second for a given connection, one timeout has a larger impact. Increasing the number of parallel connections reverts this effect as expected, as the impact of one connection experiencing a timeout is reduced.

SYN cookies are not able to process the increased amount of HTTP traffic even for low SYN flood. This is due to the increased complexity of the state keeping, causing problems for rapid port reuse as previous state still persists.

To reduce clutter, we do not show $Auth_{cookie}$. For these strategies the slope is the same as for their respective $Auth_{full}$

counterparts, however, shifted to the left, which is correlated to the reduced maximum SYN flood that can be processed. In this case, the probability also drops for the NFP platform.

Connection latencies for the best case (no SYN flood), average case (50 % SYN flood in relation to maximum processed flood) and worst case (overload at maximum processed flood) are shown in Figure 12. For most scenarios, the median latency is between 1 and 1.4 ms, while for the no flood and 50 % cases a long-tail up to 4 ms is visible. The COTS implementations show sporadic outliers and, when reaching overload, a long-tail behavior with up to a second already for the 90th percentile. Due to the mentioned lower probability when issuing 1000 HTTP requests per second for 100 parallel connections, the median latency during overload is above one second.

The exception is the NFP, which shows latencies between 1 and 4 ms without outliers even under overload. Furthermore, the latency of the no flood scenario is worse compared to when increasing the SYN flood load. We attribute this to specifics of the architecture, improving processing when increasing the backpressure on internal buffers or when reducing idle cycles caused by energy saving features.

For the $Auth_{full}$ implementation on the NetFPGA SUME the latency for up to 1000 HTTP requests per second was stable between 1 to 4 ms without outliers up to a SYN flood of 1.5 Mpps. However, for higher loads of mixed traffic the program stopped working, which we attribute to timing issues of the generated program. Related work shows that the NetFPGA SUME in general is able to run P4 programs of higher complexity with latency below 10 μs and no long-tail. This is even possible when modifying the P4 architecture of the NetFPGA to integrate a SipHash or SHA3 core, which can be used to hash even complete packet data. [46]

### F. Resource Consumption

For the NetFPGA SUME the synthesized P4 proxy program only uses up to one third of the total resources (Table III). This leaves enough resources to further enhance the program to defend against other attacks or to add a SipHash implementation into the P4 pipeline, allowing to even perform $Auth_{cookie}$.

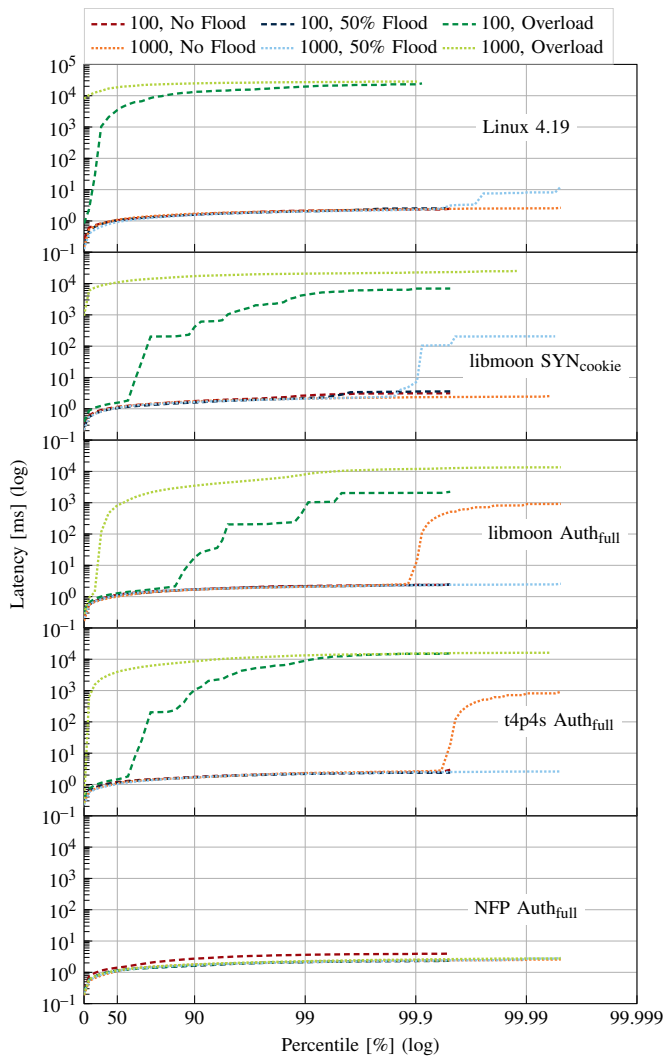

Fig. 12: High Dynamic Range latency histogram for 100 and 1000 HTTP requests per second

|  | LUTs | | Registers | | BRAM | |
|---|---|---|---|---|---|---|
|  | Abs. | % | Abs. | % | Abs. [kB] | % |
| Auth$_{full}$ | 98,343 | 22.71 | 156,678 | 18.08 | 18.450 | 34.87 |
| taken from [46] | | | | | | |
| Baseline | 64,533 | 14.90 | 109,783 | 12.67 | 16,362 | 30.92 |
| SipHash | 66,380 | 15.32 | 114,282 | 13.19 | 17,460 | 32.99 |
| SHA3-512 | 73,449 | 16.95 | 118,689 | 13.70 | 17,460 | 32.99 |

TABLE III: Resource utilization of proxy on NetFPGA SUME compared to P4 hash implementations taken from [46]

## VII. RELATED WORK

**SYN Cookie Implementations** Besides Linux, other major operating systems like Windows and FreeBSD employ TCP SYN cookies as preferred mitigation method, enabled by default during periods with high traffic volumes [38, 9].

SYN cookies, as part of a SYN proxy, have been implemented by multiple projects. A *SYNPROXY* module for the netfilter framework is available in Linux [37]. Managing the SYN flood, *SYNPROXY* forwards only legitimate traffic to the Linux kernel. To do so, the initial SYN segment is intercepted by netfilter, calculating a SYN cookie. Once the client finishes connection establishment with a verified cookie, the proxy sends a SYN segment to the original server destination, using the initially negotiated options. After finishing the second handshake, the proxy is only involved in sequence number and timestamp translation [37]. This approach enabled mitigation of a 2 Mpps SYN flood using only 7 % CPU utilization on a eight core system [37].

Other proposed techniques and implementations produce unsatisfying ($<<$ 1 Mpps) performance [29], break TCP or require specialized clients [31], or only mitigate attacks in particular scenarios, thus interfering with deployment [32].

**High-Performance Attack Mitigation in Software** As standard network stack implementations are not capable of high-performance packet processing, dedicated frameworks have been created over recent years [10, 15, 24, 45, 22]. With kernel bypass, memory pre-allocation, parallelization and batch processing, among others, deficiencies of common bottlenecks are negated, enabling packet processing beyond 10 Gbit/s on commodity hardware.

DDoStop [34] is an application based on the Snabb virtual switch [25], a framework for fast packet networking. This NFV-like app offers zombie detection, a term coined by Arbor Networks, i.e., blocking of "source hosts that exceed certain thresholds" [34]. Subject to the complexity of the rule-set and required packet parsing, 1 Mpps to 10 Mpps of general DoS traffic, per core, were mitigateable [34].

## VIII. CONCLUSION

SYN flooding is the predominant traffic for high-volume (D)DoS attacks on the Internet and will likely remain so in the future as the root-cause cannot be fixed. SYN cookie and SYN authentication can prevent memory exhaustion caused by SYN floods at the price of increased computational complexity. Our case study demonstrates that computational costs are too high to effectively defend against SYN flood attacks on the end host itself. However, the client puzzle—including a cryptographic hash value—is currently the single effective defense strategy, guaranteeing that no malicious connection attempts are successful.

A more scalable solution is a SYN proxy running on a dedicated node. This allows the protection of entire networks without increased resource usage. Our prototype implementations based on libmoon, as well as the P4 language used to program software and hardware data planes, show that SYN cookie and SYN authentication, when used in a proxy setup, scale to mitigate SYN flooding at 10 GbE line-rate. The libmoon proxy scales well and allows almost arbitrary complex packet processing, while the P4 solutions are easier to implement and can be ported to different target platforms, in particular hardware devices, which achieve lower latency with less or no outliers.

We conclude that effective and efficient SYN flood mititation on modern data planes is possible. With both migitation strategies performing equally, we recommend SYN authentication, being the simpler to implement. The crucial limiting factor for hardware data plane solutions is the availability of a suitable cryptographic hash function. However, cryptographic hash operations can be implemented in hardware efficiently—demonstrated by large-scale experimental prototypes such as Bitcoin—which would allow for powerful data plane driven SYN flood migitation.

## Availability

As part of our ongoing effort towards repeatable, replicable and reproducible network research, we release the source code of our SYN proxy implementations for libmoon and P4 targets at https://github.com/syn-proxy/implementations.